%% file: IEEE.tex
\documentclass[conference]{IEEEtran}

\input{assets/documents/00packages.tex}
\usepackage[numbers]{natbib}
\let\oldcitep\citep
\renewcommand{\citep}[1]{\mbox{\oldcitep{#1}}}
\let\oldcitet\citet
\renewcommand{\citet}[1]{\mbox{\oldcitet{#1}}}
\let\oldcite\cite
\renewcommand{\cite}[1]{\mbox{\oldcite{#1}}}

\begin{document}

\title{Generalised Rate Control Approach For Stream Processing Applications}

\author{\IEEEauthorblockN{Ziren Xiao}
\IEEEauthorblockA{
\textit{The University of Melbourne}\\
Melbourne, Australia \\
zirenx@student.unimelb.edu.au}
}

\maketitle

\input{assets/documents/01abstract.tex}


\IEEEpeerreviewmaketitle

\input{assets/documents/01_Introduction}

\input{assets/documents/02_Related_work}

\input{assets/documents/03_System_model}

\input{assets/documents/04_Method}

\input{assets/documents/05_Experiment}

\input{assets/documents/06_Conclusion}

\section*{Acknowledgement}
This research was supported by the China Scholarship Council.

\bibliographystyle{IEEEtranN}
\bibliography{references.bib}

\end{document}

%% file: assets/documents/00packages.tex
\usepackage{lineno,hyperref}
\usepackage{amsmath,amssymb,amsfonts}
\usepackage{graphicx}
\usepackage{textcomp}
\usepackage{xcolor}
\usepackage{algorithm}
\usepackage{algorithmic}
\usepackage{wrapfig}
\usepackage{booktabs}
\usepackage{multirow}
\usepackage{multicol}
\usepackage{subcaption}
\usepackage[final]{changes} 

%% file: assets/documents/01abstract.tex
\begin{abstract}
Distributed stream processing systems are widely deployed to process real-time data generated by various devices, such as sensors and software systems. A key challenge in the system is overloading, which leads to an unstable system status and consumes additional system resources. In this paper, we use a graph neural network-based deep reinforcement learning to collaboratively control the data emission rate at which the data is generated in the stream source to proactively avoid overloading scenarios. Instead of using a traditional multi-layer perceptron-styled network to control the rate, the graph neural network is used to process system metrics collected from the stream processing engine. Consequently, the learning agent (i) avoids storing past states where previous actions may affect the current state, (ii) is without waiting a long interval until the current action has been fully effective and reflected in the system's specific metrics, and more importantly, (iii) is able to adapt multiple stream applications in multiple scenarios. We deploy the rate control approach on three applications, and the experimental results demonstrate that the throughput and end-to-end latency are improved by up to 13.5\% and 30\%, respectively.
\end{abstract}

%% file: assets/documents/01_Introduction.tex
\section{Introduction}
\label{Introduction}

The widespread adoption of the Internet along with an ever-growing number of interconnected devices and an increasingly digitally-reliant society, has led to vast amounts of data being generated at unprecedented rates, a phenomenon known as big data. This data is not only large in volume but also exhibits complexity and heterogeneity in terms of semantics and structure, taking, for example, the form of text, images, audio, and video. This poses a significant challenge - processing streaming, or continuously generated data, in real-time efficiently with the aim of providing value-adding and timely insights to individuals and organisations. Existing frameworks, namely \emph{Distributed Stream Processing Systems} (DSPSs), rely on distributed computing paradigms and platforms, such as clusters in cloud data centres, to aggregate compute and storage power to meet the resource and quality of service demands of real-time streaming applications. DSPSs offer real-time platforms to responsively process continuously incoming data from external sources, such as Internet of Things (IoT) sensors, cameras on intelligent vehicles and online information crawlers. 


In general, the majority of methods used to optimise the performance of a DSPS include operator scheduling (i.e., scaling and placement) and parameter tuning. The solved problems can be an NP-hard problem~\cite{thiele2000real,cardellini2016optimal} or the solution may be designed and characterised for a specific problem~\cite{nardelli2019efficient}. Alternatively, a rate control method in \cite{xiao2024intelligent} is proposed to dynamically throttle the emission rate at the \emph{source} component to avoid entering an overloading state. However, this method exposes three key shortcomings. First, the interval of collecting metrics from the system needs to be manually adjusted through different topologies. This is because the training agent must wait a certain amount of time after changing the emission rate, until the new rate has been effective in all components in the topology due to flushing queuing tuples. A long waiting time leads to an intolerable training time for the DRL agent, while a short time results in the wrong prediction made by the DRL agent because the latest status is not reflected in the system metrics. Second, states of selected system metrics are stacked with a historical size because the current status in the sink component may be a result of previous actions. Again, this historical size is a parameter that must be configured appropriately before deploying the desired topology. The configuration of both parameters requires prior knowledge about the topology, such as parameter study, which violates automatic rate control. Third, the trained model can only fit the topology we used to learn patterns, where the model has to be retrained when a new topology is deployed on the system.

Therefore, in this paper, we use a graph neural network-integrated proximal policy optimisation (GPPO) proposed in \cite{xiao2024macns}, which was originally used to address dynamic multi-agent navigation problems, to overcome those shortcomings. 
GPPO demonstrates two competitive advantages in addressing these shortcomings against other Multi-Layer Perceptron (MLP)--based DRL algorithms. First, GPPO accepts graph data as the input, which perfectly matches a graph-based topology. That is, the DRL agent knows structured data from each component of the topology and finds the patterns, connections and correlations of system metrics between components. This ensures the DRL agent can capture all metrics from the system, while the MLP-based agent must filter key features (e.g., perform feature engineering) to accelerate the training. This is the key feature that the DRL agent can make a collaborative decision based on all components in the system, rather than only considering metrics from the \emph{source} and \emph{sink} in \cite{xiao2024intelligent}. Consequently, the stack of states can be waived and we can set a small and fixed time on the interval of collecting metrics. Second, GPPO provides a new training strategy to the environment that can accept the dynamic size of states. In other words, the state space can be dynamically changing without fully retraining the DRL agent, if the shape of the input topology was never seen before. This feature offers a new training mode to the DRL agent, enabling learning patterns or making predictions from multiple different topologies simultaneously.

The proposed method proactively limits the data emission rate at the streaming source to achieve a higher system throughput and a smaller end-to-end latency caused by the system's overloading mechanisms. To control the rate via a graph-based DRL agent, the state of each component in the DSPS is merged and converted to a Directed Acyclic Graph (DAG). 
A vertex in the graph represents a stream component in the system, which involves multiple vertex features (i.e., the metrics on the component), such as the data arrival rate, processing latency, queuing latency and send rate. Vertices are connected by directed edges showing the data flow direction. 
The transformed DAG is then forwarded to GPPO, which returns a discredited action indicating the level of throttling to be limited at the stream source. Through our experiments, we demonstrate our approach achieves superior throughput and latency than the back pressure mechanism in different stream processing applications, and also shows the possibility of controlling different topologies via the same DRL agent without fully retraining the model. The experimental results demonstrate our proposed approach is able to adapt multiple topologies in multiple different scenarios, and the performance is superior to the system's default scheme using back pressure only.

The main contributions of this paper are summarised as follows:
\begin{itemize}
    \item We model the dynamic rate control mechanism as a Directed Acyclic Graph (DAG), which can be further formulated as a graph-based reinforcement learning problem. This shows the possibility of learning the acquired knowledge across different topologies without fully retraining the DRL model.
    \item The proposed method is parameter-free compared with the previous method, such as the historical size and metrics collecting interval. This is especially useful when there is a lack of exact knowledge of the running topology, which is the actual situation faced in real distributed stream processing systems.
    \item The well-trained model can be deployed on the stream application, which does not require changes in the system's source code.
\end{itemize}


%% file: assets/documents/02_Related_work.tex
\section{Related work}
\label{Related work}

Existing approaches to deal with the overloading problem mainly focus on back pressure, load shedding, operator placement, operator scaling and system parameter tuning. The back pressure is a flow control strategy aiming to regulate the emission rate from upstream, starting at where the overloading occurs. The back pressure implementation in Apache Storm \cite{apachestorm} relies on the capacity of the incoming queues. Specifically, when the number of tuples in the queue exceeds its maximum capacity, the back pressure status signal is multicasted to its upstream operators or source. Once those components have received the signal, they temporarily stop emitting tuples until the resuming signal is received i.e., the queue's length is below its capacity.

Recently, the integration of multiple optimisation techniques into reinforcement learning (RL) has been widely considered in many papers. One of the main ideas of those combinatorial optimisations in DSPS is to use graph representation models. \citet{addanki2019placeto} provide a task graph mapping approach with improved generalisation using graph embedding techniques. A similar approach is proposed in \cite{zhou2019gdp}, namely GDP, with a better performance. Both two approaches provide transfer learning feasibility with a constraint of the running environment where the target platform must be the same as the one used to perform training. 
However, turning a DAG representation into a set of allocating computational resources while satisfying different performance objectives, such as high throughput and low latency, can be known as an NP-hard problem in the strong sense \cite{garey1979computers}. 

Therefore, instead of relying on a hard-coded pre-processing of the DAG into a look-up table, \citet{huang2021tata} introduces a novel method, namely TATA, for optimising task placement in DSP environments, which uses a resource-aware DRL framework using graph embedding and an attention mechanism for efficient task scheduling. By using GNNs, the system can model both task and resource structures as graphs, where nodes represent tasks and resource slots. The system can capture complex task dependencies and resource relationships through graph convolution, allowing for efficient encoding of both upstream and downstream data flows. READYS \cite{grinsztajn2021readys} leverages Graph Convolutional Networks (GCNs) to dynamically schedule the task graphs and achieve better scalability and generalisation. It allows transferring a learned model to other task graphs with different shapes of the DAGs.

Although graph neural networks have been used in addressing task placement problems, overall, there is no existing approach that can dynamically throttle the data emission rate at the source component to cope with different task sizes (e.g., different parallelism of operators) and allow the transfer of the learned model to other topology settings.

%% file: assets/documents/03_System_model.tex
\section{Data Emission Rate Control Problem in DSPS}
\label{system model}

In a DSPS, the data processing architecture is strategically organized into a topology comprising sources, operators, and sinks, each fulfilling distinct roles within the data stream. Source nodes act as the entry points for the system, capturing data from a diverse array of external sources such as IoT sensors and social media platforms. These nodes initiate the flow of data through the DSPS topology. As the data progresses through the system, it is handled by operator nodes that are tasked with executing various transformations, including filtering, aggregation, or more complex data manipulations. These operations refine the data as it traverses the network. Concluding the processing journey, the data reaches sink nodes, which serve as the termination points in the topology. Here, data is consolidated, stored, and possibly prepared for transmission to other systems or for display purposes. This structured topology ensures a seamless progression from source to sink, optimizing each segment of the network to efficiently manage the flow and processing of data, thereby maintaining the integrity and efficiency of the entire system. The \textbf{Rate Control} mechanism is designed to optimize data flow dynamically across the network, ensuring that the system operates within its capacity limits while maintaining high performance in terms of throughput and latency. The central component of the rate control mechanism is dynamic rate adjustment, which is implemented to modulate the data emission rates at source nodes based on current system conditions. This process involves a feedback loop where data processing nodes continuously report their status to the central coordinator. Key metrics include the current load, processing speed, and queue lengths. Based on these metrics, the central coordinator calculates the optimal emission rates to prevent underutilization and congestion in the network. 



The management of the data emission rate at these source nodes is critical. Without effective control, the emission rate could exceed the processing capabilities of the subsequent nodes in the system, leading to data backlogs and increased latency. The dynamic mechanism of rate control based on real-time feedback about system performance and current processing capacities is a good solution. Therefore, the primary objectives of controlling the data emission rate include maximising throughput to ensure the system processes the maximum number of data units efficiently without congestion, minimising latency to reduce the total time from data entry to exit, and balancing the workload across nodes to prevent any node from becoming a bottleneck. Achieving these goals is crucial for maintaining efficient system operation under varying data loads, ensuring that the DSPS provides reliable data processing and supports timely decision-making.

In our study, we abstract DSPS topologies as directed graphs where vertices represent data processing nodes—sources (\(src\)), operators (\(op\)), and sinks (\(sk\))—and edges denote the data streams that interconnect these nodes. Source nodes, characterized by their data generation rate (\(r_g\)) and absence of incoming edges, initiate data flow and are additionally detailed by their current data emission rate (\(r_c\)), the total time in back pressure status (\(\text{src}_{bk}\)), the number of tuples emitted from the outgoing queue (\(\text{src}_{out}\)), and the maximum queue capacity of the outgoing queue (\(\text{src}_{max}\)). Operators, positioned between sources and sinks, manage data transformations and are defined by metrics such as the number of tuples accepted from upstream (\(\text{op}^{i}_{in}\)), emitted from the outgoing queue (\(\text{op}^{o}_{out}\)), and the capacities of incoming (\(\text{op}^{i}_{max}\)) and outgoing (\(\text{op}^{o}_{max}\)) queues, along with the total time spent in back pressure status (\(\text{op}_{bk}\)). Sinks, as the endpoints, finalize data processing and are characterized by the number of tuples accepted from upstream (\(\text{sk}^{i}_{in}\)), the maximum queue capacity of the incoming queue (\(\text{sk}^{i}_{max}\)), the sum of latency from source to sink (\(\text{sk}_l\)), and the number of tuples processed (\(\text{sk}_p\)). This topology is encapsulated by a Directed Acyclic Graph (DAG) \(G = (V, E)\), where \(V\) and \(E\) represent the nodes and their connecting data flows, respectively. Our objective is to optimize the emission rate (\(r_s^*\)) at the source to maximize throughput without triggering back pressure mechanisms, thereby ensuring efficient operation under varying system loads. We define the calculation of throughput (\text{thr}) over the last \(K\) seconds. Throughput is calculated as the sum of the number of tuples processed at the sink nodes ($\text{sk}_p$) over \(K\). This metric is given by the formula:
\[
\text{thr} = \frac{\sum^{K} \text{sk}_p}{K}
\]



%% file: assets/documents/04_Method.tex
\section{Collaborative Rate Control with DRL}
\label{System Design}
\begin{figure*}
    \centering
    \includegraphics[width=1\textwidth]{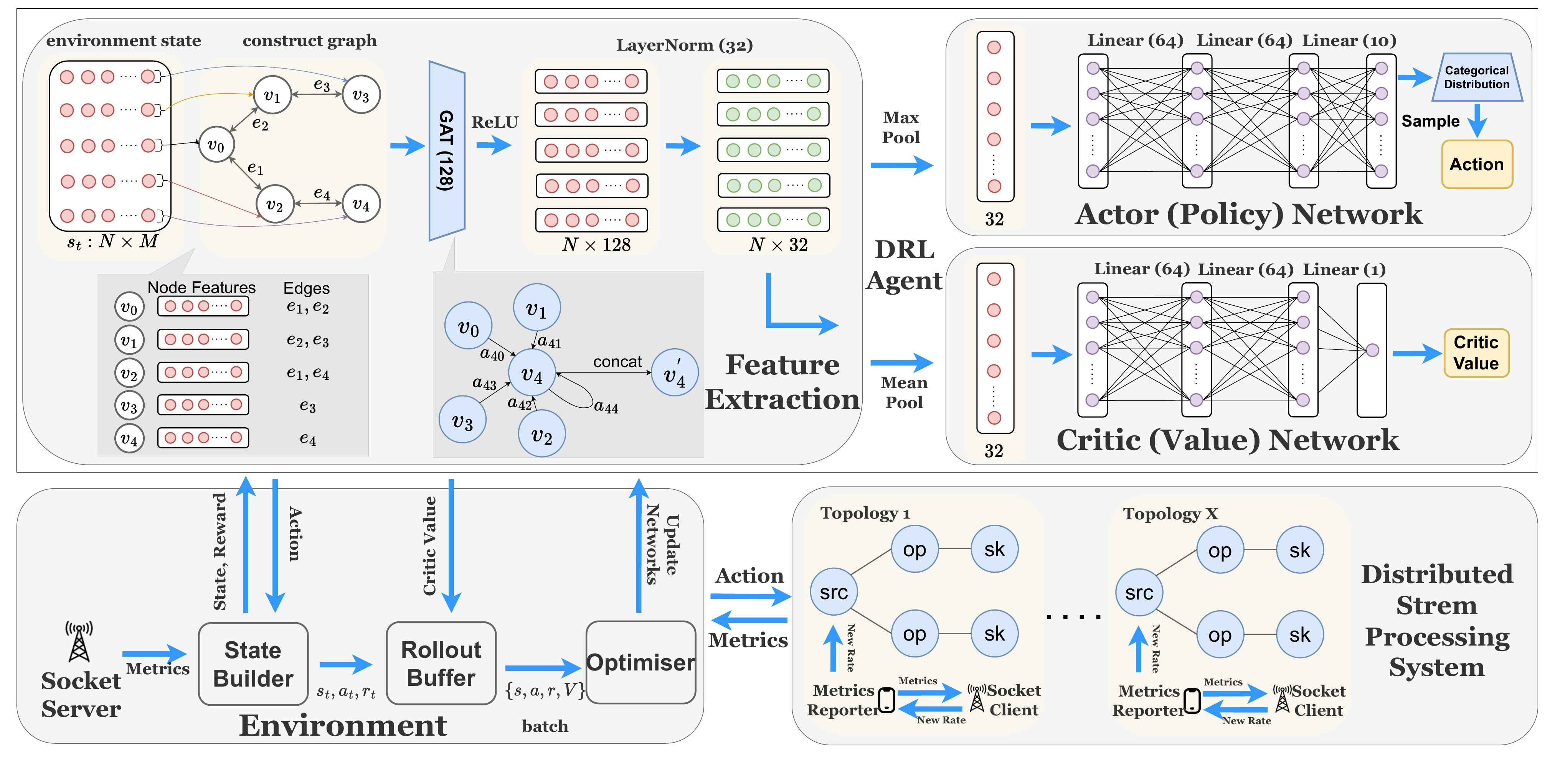}
\caption{The framework of our DRL-based Rate Control Approach}
\label{fig:workflow}
\end{figure*}

We use the idea from Graph-based Proximal Policy Optimisation (GPPO) \cite{xiao2024macns} to build our rate control framework, as shown in Fig.~\ref{fig:workflow}. To better fit our specific problem, we modify the structure of GPPO, where Graph Neural Networks (GNNs) extract key features and pass them to both actor and critic networks. Specifically, the Metrics Reporter collects system metrics of the topology (application) in the DSPS runs for the last $K$ seconds and reports the collected data to the environment. The environment builds a DRL agent-recognisable state and passes it to the Feature Extractor in the DRL agent. The extracted features are further forwarded to the actor and critic network via Max Pool and Mean Pool functions, respectively. The actor network responds an action to the environment, which is then sent back to the topology to adjust its emission rate. The critic network returns a critic value and stores it in the Rollout Buffer along with the corresponding state, action and reward. The Optimiser fetches data from the Rollout Buffer periodically and updates the Feature Extractor, actor and critic neural networks. In the following subsections, we describe each component in the framework specifically. 

\subsection{Environment}
Our framework is pluggable into different DSPS environments and DSP applications without changing the system's source code via communicating with the environment component. The environment is built atop a standard \emph{OpenAI Gym} interface. We formulate the problem as a Markov Decision Process (MDP): ($S$, $A$, $R$, $P$), where the transition probability $P$ is controlled by the neural networks instead of a fixed value. 

\paragraph{State Space $S$}
Once the environment receives system metrics from the DSPS, it passes those values to a State Builder, which transforms metrics into the state $s_t$ at the specific time step $t$. Each $s_t$ consists of metrics from all components in a DSP application during the past $K$ seconds.
For the \emph{source} component, it involves (i) the maximum available data emission rate (i.e., the data generation rate \(r_g\)), (ii) the current data emission rate \((r_c)\), (iii) the total time in the back pressure status in the interval \((\text{src}_{bk})\), (iv) the number of tuples emitted from the outgoing queue \((\text{src}_{out})\), and (v) the maximum queue capacity of the outgoing queue \((\text{src}_{max})\). Each \emph{operator} component includes (i) the number of tuples accepted from the upstream \((\text{op}^{i}_{in})\), (ii) the number of tuples emitted from the outgoing queue \((\text{op}^{o}_{out})\), (iii) the maximum queue capacity of the outgoing queue \((\text{op}^{o}_{max})\), (iv) the maximum queue capacity of the incoming queue \((\text{op}^{i}_{max})\), and (v) the total time in the back pressure status \((\text{op}_{bk})\). The \emph{sink} component reports metrics of (i) the total time in the back pressure status \((\text{sk}_{bk})\), (ii) the number of tuples accepted from the upstream \((\text{sk}^{i}_{in})\), (iii) the maximum queue capacity of the incoming queue \((\text{sk}^{i}_{max})\), (iv) the sum of the latency from the source to sink \((\text{sk}_l)\), and (v) the number of tuples processed \((\text{sk}_p)\).
Unlike the approach in \cite{xiao2024intelligent}, we have omitted the historical metrics because the state collaboratively reflects the overall status of the system, while \cite{xiao2024intelligent} only focuses on a few specific key features - the performance of the previous approach can be largely affected by metrics that indicate weak or no correlations to the emission rate.

\paragraph{Action Space $A$}
The DRL agent's action space is modelled in terms of the data generation rate $r_g$, the maximum available data emission rate from the \emph{source} component to downstream. The action $a_t$ at time $t$ is from the discredited action space $A=\{0.1,0.2,0.3,0.4,0.5,0.6,0.7,0.8,0.9,1.0\}$, indicating a fraction of $r_g$. Specifically, the current emission rate at time $t$ is set to $a_t \times r_t$ with an action $a_t \in A$.

\paragraph{Reward Function $R$}
We observe that the system shows the maximum throughput and minimum end-to-end latency when the data emission rate is around (but less than) the threshold of entering the back pressure status, which perfectly fits our objectives: maximising the application's throughput and minimising the latency. These objectives must be reflected in the reward function design. Therefore, we use the min-max normalised throughput as the reward at time $t$:

\[r_t = R(s_t, a_t) = \frac{thr - thr_{min}}{thr_{max} - thr_{min}}\]
where $thr$ is the current throughput during the last $K$ seconds, $thr_{max}$ and $thr_{min}$ are the maximum and minimum throughput that the agent observed in the past, respectively. The main purposes of using a normalised reward are (i) the gap between the optimal throughput (i.e., the throughput of the emission rate at the threshold of triggering back pressure) and the throughput of the rate without any limitation is not significant, while the min-max normalisation can magnify the difference; (ii) training across different topologies becomes easier because the maximum possible reward of each time step is 1, whereas the non-normalised throughput can be largely different. 

\subsection{DRL Agent}

\paragraph{Graph Construction and Feature Extraction}
In this approach, we use GNNs as the feature extractor because GNNs are particularly well-suited for feature extraction in DSPS due to their inherent ability to capture complex dependencies and interactions within graph-based data structures, especially for DAGs. Specifically, the state from the State Builder is first transformed into a DAG with the same representation as in the topology. The node features involve metrics from the corresponding stream component, and the edge features indicate the link information (e.g., maximum link bandwidth and link latency). The constructed graph is then forwarded to the graph neural network, Graph Attention Networks (GAT), with a ReLU activation function. After a linear layer, normalisation is performed to make the distribution of the features more consistent. The result again passes through a linear layer and is forwarded to actor and critic networks. By applying GNNs as the feature extractor, the key information from observations (i.e., the states) with different sizes is extracted, enabling the DRL agent to learn or adapt from multiple different topologies. More importantly, the key information is automatically selected by the GNNs instead of manually choosing. This significantly reduces the time spent on parameter study and feature engineering. 

\paragraph{Actor (Policy) and Critic (Value) Networks}
We use a three-layer Multi-Layer Perceptron (MLP) neural network for both the actor and critic function approximation, referred from the default settings in \emph{stable baselines 3}~\cite{pposb3}, with 64 neurons in the first two layers and 10 neurons and 1 neuron in the last layer, respectively, using the Tanh as the activation function after each layer.

\subsection{Learning Process}
The Rollout Buffer records and stores each $K$ seconds interaction between the environment and topology. At every 2,048 time steps (one iteration), the optimiser fetches a set of $(s,a,r,V)$ from those steps and splits into 32 batches (with 64 samples in each batch) to update the feature extractor, actor, and critic networks. We use the same PPO parameter settings from \cite{stable-baselines3} to train the DRL agent: $\lambda=0.99, \lambda_{gae}=0.95, clip=0.2, entropy\_coef=0, value\_coef=0.5$ and $learning\_rate=0.0003$.

%% file: assets/documents/05_Experiment.tex
\section{Experiment}
\label{Performance Evaluation}

In this section, we report experimental results in which we compare our GNN-based rate control approach with the system's default scheme.

\subsection{Experiment Setup}

\subsubsection{Environment}
We implement the experimental environment under a network simulator, OMNet++. In particular, the simulator is used to simulate the workflow in the stream processing system and is able to run streaming applications. In all simulated experiments, application operators are deployed on virtual cloud servers, interconnected by a 100Mbps in/out Ethernet port with an average delay of 0.5ms. Each \emph{operator} is associated with an incoming queue, which accepts and temporarily stores data (i.e., tuples) from upstream, and an outgoing queue, which stores pending emitted data to downstream. While the \emph{source} and \emph{sink} components only contain an outgoing queue or an incoming queue, respectively. The maximum capacity of all incoming and outgoing queues is set to 64. To simulate the adaptability of the data emission rate, a multiplier of a random range [0.7, 1.3] is applied to the original rate at every 100 time steps during the training, indicating a small fluctuation in the rate.

\subsubsection{Back Pressure}
We simulate the back pressure mechanism based on the implementation in Apache Storm. When a queue length exceeds its maximum capacity, the associated \emph{operator} is considered as under back pressure. The back pressure message (either entering or exiting back pressure) is sent to a Nimbus, which acts as the master node, and the message is then delivered to all upstream operators to suspend emitting. All suspended operators check the latest back pressure status every 0.1 ms (the default value in Apache Storm \cite{stormperformance}) rather than performing a busy wait. Once the \emph{operator} exits the back pressure (i.e., the number of elements in the queue is less than the maximum capacity), a message is sent out again to the Nimbus, allowing upstream operators to resume the data emission tasks. In our simulation environment, we delay the currently executing job due to emitting the back pressure message with 0.05ms, which is an average value profiled in the real-world deployment application in Apache Storm. 

\subsection{DSP Applications}
\begin{figure}
    \centering
    \includegraphics[width=0.5\textwidth]{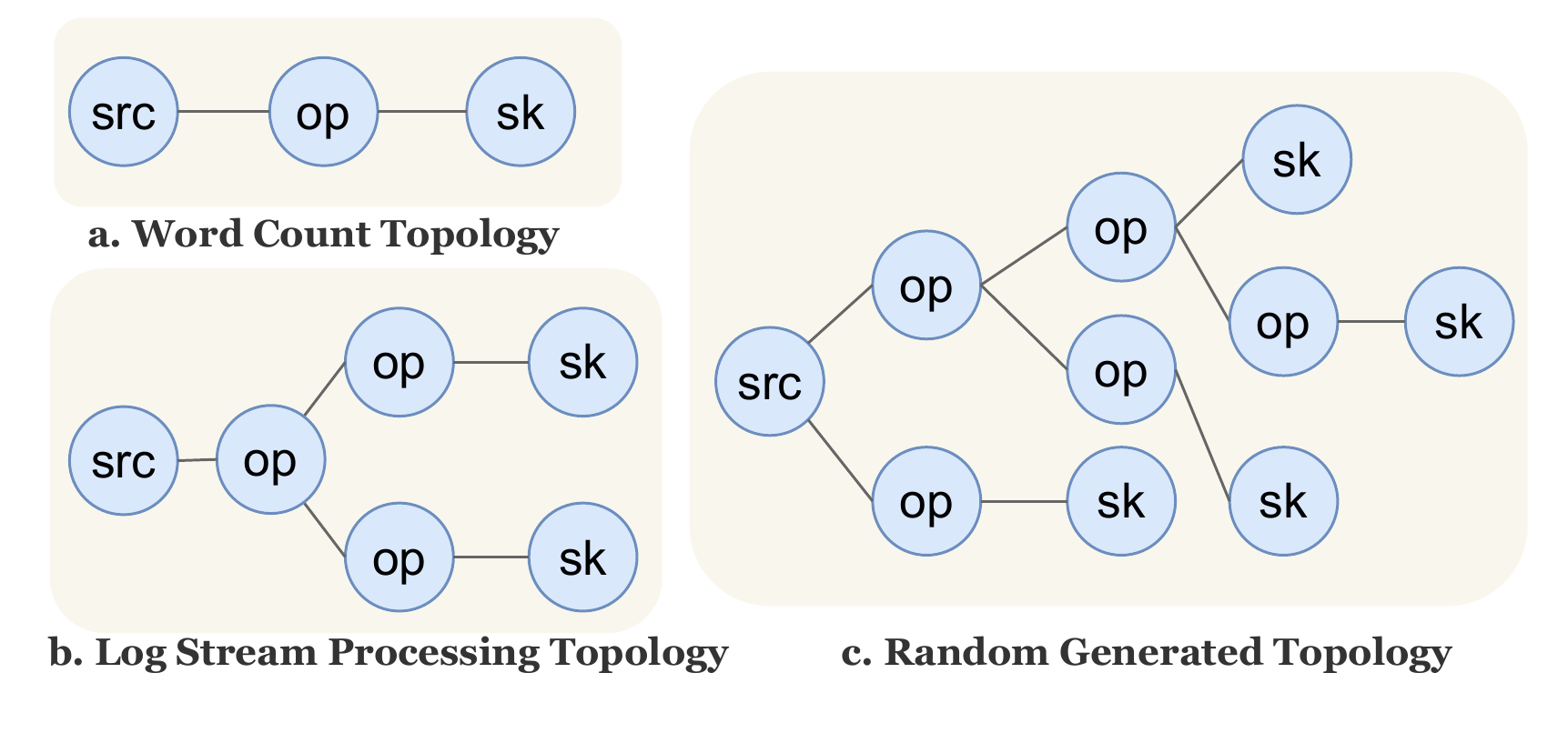}
\caption{Graph representations of the evaluated distributed stream processing applications (topologies). src, op and sk represent \emph{source}, \emph{operator} and \emph{sink} stream component, respectively.}
\label{fig:topology}
\end{figure}

We consider three topologies to evaluate the performance, including the word count topology and log stream processing topology used in real-life deployment and randomly generated complex topology involving 10 individual stream components. The details are illustrated in Fig.~\ref{fig:topology} and described below:

\begin{itemize}
    \item Word Count Topology (WCT): The word count topology is a well-known stream application used to count the number of appearances for each individual word in a file. It involves three stream components: (i) the \emph{source} component emits a sentence consisting of a random number of words; (ii) the \emph{split} operator splits a sentence into words, and (iii) the \emph{count} operator records the corresponding word's appearances.
    \item Log Stream Processing Topology (LSPT): The log stream processing topology is a real-world use case that stores and analyses logs. It consists of six stream components: (i) a random-sized log entry is generated and emitted at the \emph{source} component to simulate the real-world case, (ii) the \emph{rule} operator analyses the log entry based on specific rules and emit it to both \emph{indexing} and \emph{counting} operator, (iii) after \emph{indexing} and \emph{counting} operators, the result is emitted to two separate \emph{sink} bolt.
    \item Randomly Generated Topology (RGT): We randomly generate a tree-styled complex topology to evaluate the performance of our approach under multi-level applications. 
\end{itemize}

Both parameter settings from WCT and LSPT, such as the processing latency and the maximum data generation rate, are profiled by real topology implementations in Apache Storm.

\subsection{Experiment Results}

\subsubsection{Multiple Topology Adaptability}
\begin{figure}
 \centering
 \includegraphics[width=0.5\textwidth]{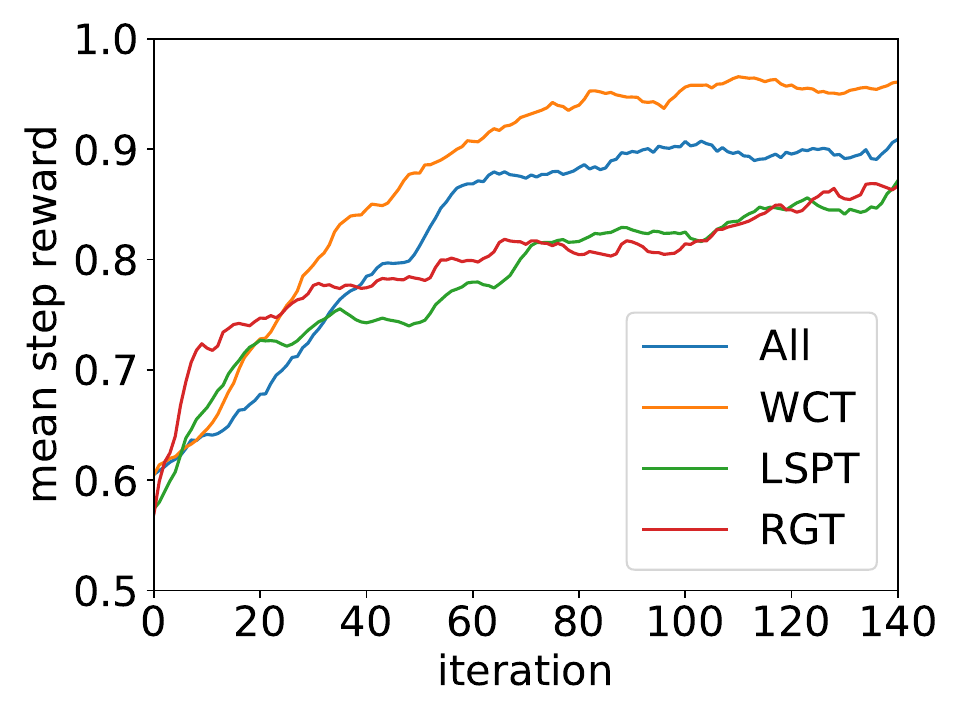}
 \caption{All-in-one model training performance.}
 \label{fig:aio-train}
\end{figure}

The main advantage of using graph neural networks is the adaptability of multiple topologies. Specifically, the well-trained DRL agent should be able to fit multiple stream applications rather than being limited to only one specific topology or one particular problem. Furthermore, the DRL agent should be able to learn new topologies or patterns that had not yet been seen in the previous training. To achieve these objectives, all three topologies are trained with the same DRL agent at the same time using a distributed training strategy. 

Fig.~\ref{fig:aio-train} shows the DRL agent training progress of all three topologies at the same time (labelled as `All', we name it as `all-in-one' in this section). At each iteration, the environment collects an equal number of metrics from those three topologies (i.e., equivalently to $2048 / 3$ time steps per topology per iteration). As can be seen from the graph, the training performance of the all-in-one model is in between three models with individual training. That is, the LSPT and RGT can benefit from this combined training because of shorter training time and faster convergence, or otherwise they both require further training iterations to approach the optimal solution. The main reason can be the training of WCT is easier than the other two topologies due to the topology complexity, while the all-in-one model can still learn patterns from WCT and contribute to understanding others. 

Note that even though our all-in-one model has achieved around 0.92 reward out of 1 (maximum theory value due to min-max normalisation), there is a small gap to reach the optimal mean step reward. This is because (i) the PPO algorithm is a stochastic algorithm, which includes both exploration and exploitation during the training (whereas the test result is deterministic). As a result, there is a small chance that the DRL agent selects non-optimal action to explore more possible states, which causes a drop in the mean reward; (ii) the maximum reward used to calculate the normalisation may indicate the best luck in the randomness in the simulation, such as the minimum network latency and the fastest processing time, which can rarely be reproducible. We analyse and demonstrate the test results in the later subsection to validate the performance of our proposed method.

\subsubsection{Parameter Settings}
\begin{figure}[ht]
         \centering
         \includegraphics[width=0.5\textwidth]{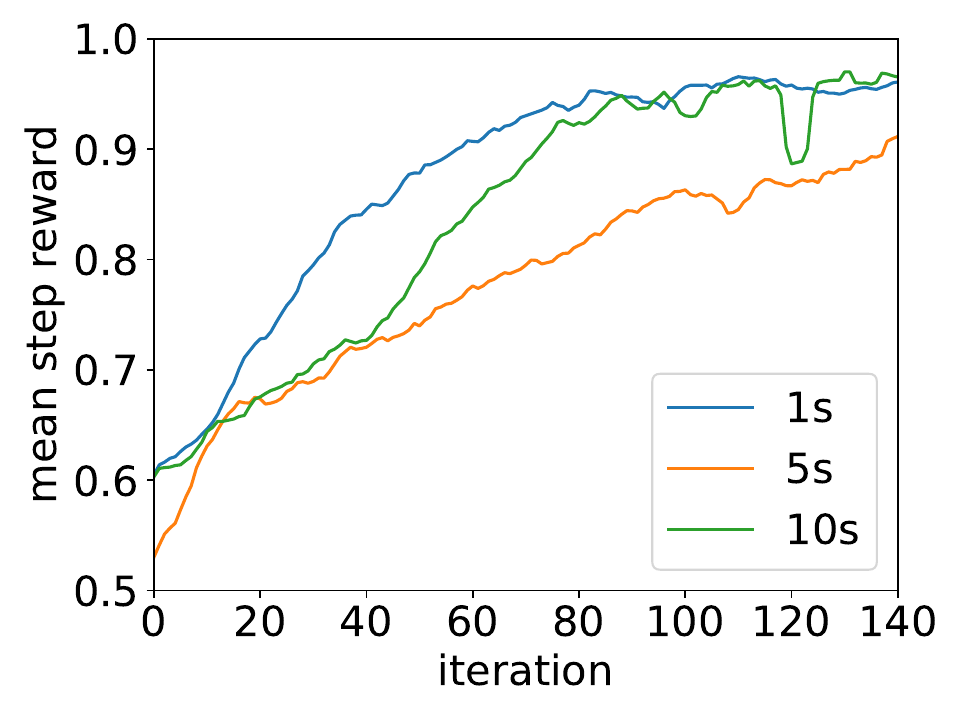}
        \caption{Parameter study of the metrics collection interval in Word Count Topology.}
        \label{fig:mci}
\end{figure}
As we have mentioned in the previous sections, the manual parameter settings of historical size and metrics collecting interval are removed in our GPPO-based approach, which was originally mandatory in \cite{xiao2024intelligent}. We compare different metrics collecting intervals in the WCT. As can be seen from Fig.~\ref{fig:mci}, the interval settings in 1s and 10s have a similar converge trending, while the one with 5s shows a slower converge speed. This is because PPO is a stochastic algorithm where the neural networks are initially randomised. That is, the initialisation of the neural networks can be `unlucky', which requires more training iterations to approach the optimal solution. However, a larger collecting interval indicates a longer training time, e.g., the 10s setting costs 10 times the training time than 1s. Therefore, a smaller interval is suggested (e.g., minimum 1s in Apache Storm) due to (i) similar training performance and (ii) faster training time.

\subsubsection{Performance Evaluation}
\begin{figure*}
     \centering
     \begin{subfigure}[b]{0.44\textwidth}
         \centering
         \includegraphics[width=\textwidth]{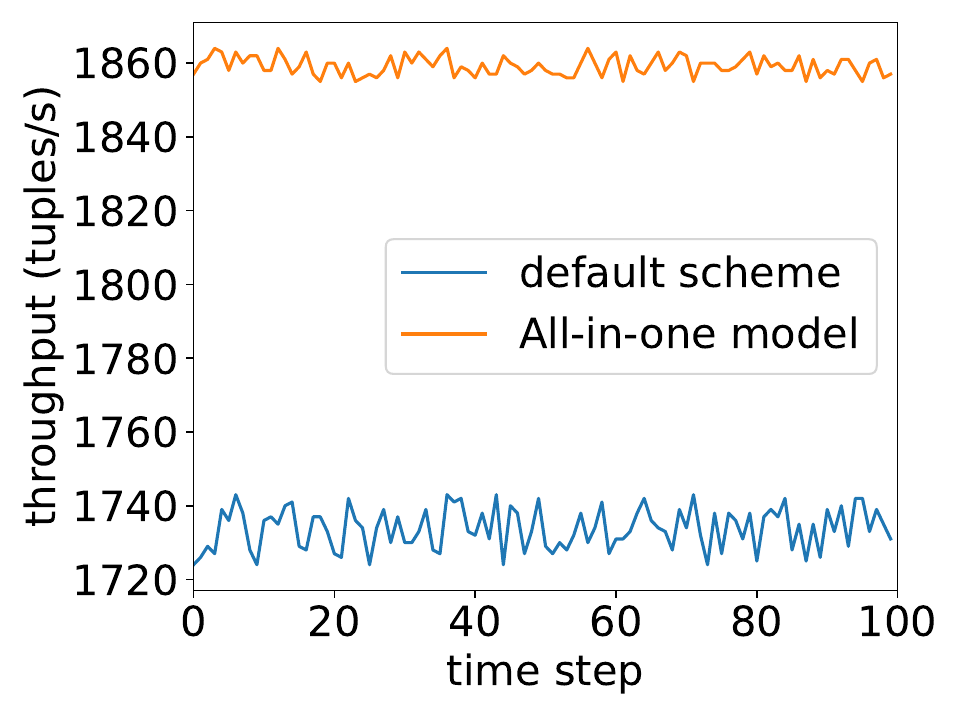}
         \caption{Throughput of WCT.}
         \label{fig:aio-test-wc}
     \end{subfigure}
     \begin{subfigure}[b]{0.44\textwidth}
         \centering
         \includegraphics[width=\textwidth]{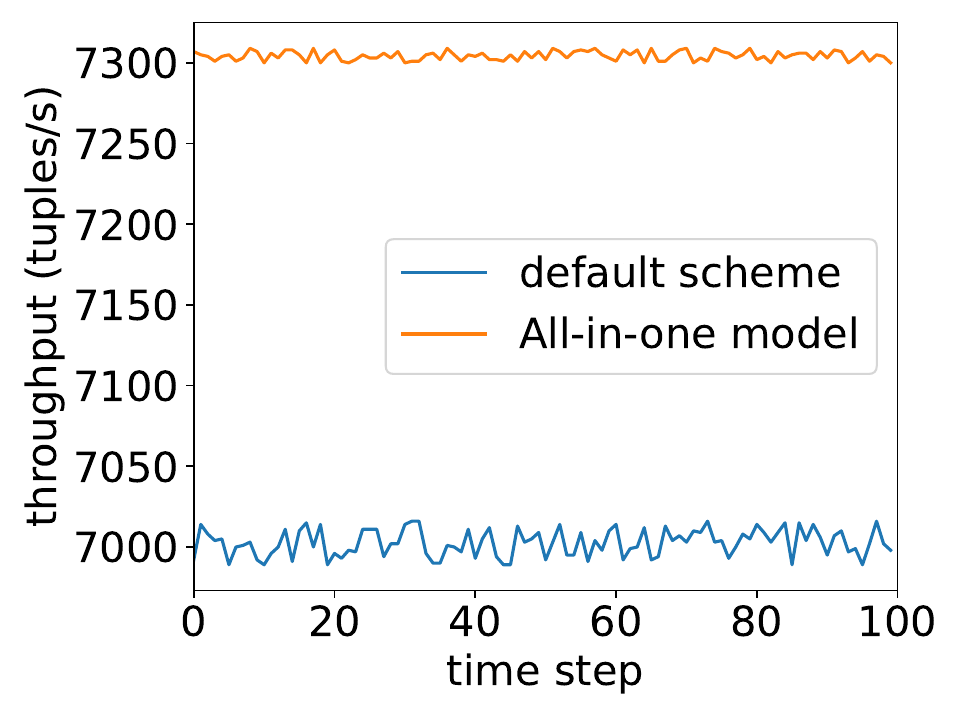}
         \caption{Throughput of LSPT.}
         \label{fig:aio-test-lsp}
     \end{subfigure}
     \begin{subfigure}[b]{0.44\textwidth}
         \centering
         \includegraphics[width=\textwidth]{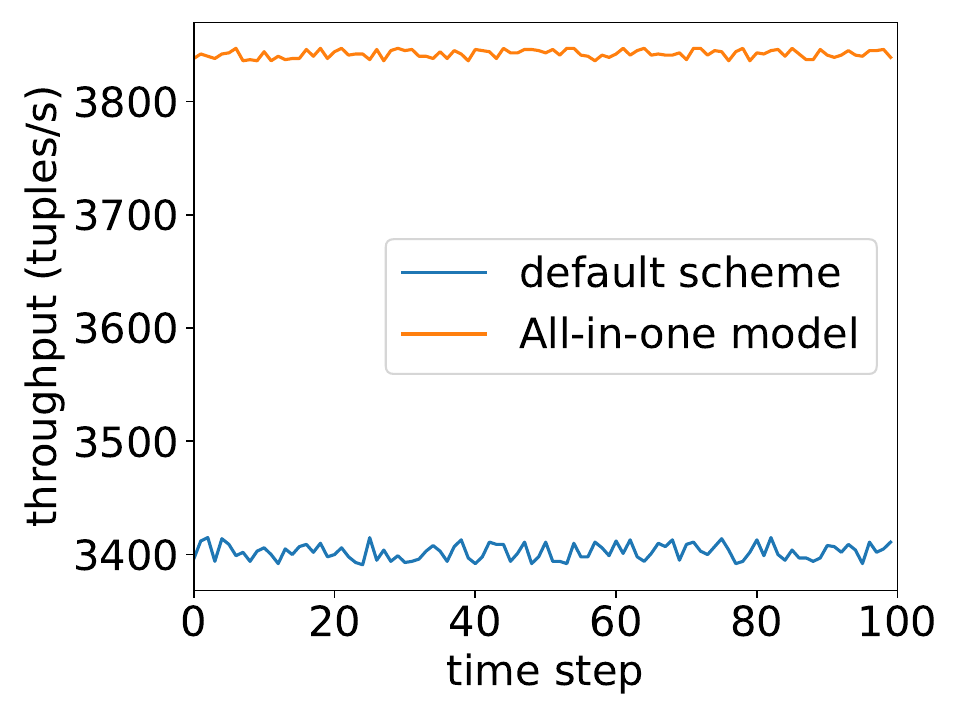}
         \caption{Throughput of RGT.}
         \label{fig:aio-test-rg}
     \end{subfigure}
    \begin{subfigure}[b]{0.44\textwidth}
         \centering
         \includegraphics[width=\textwidth]{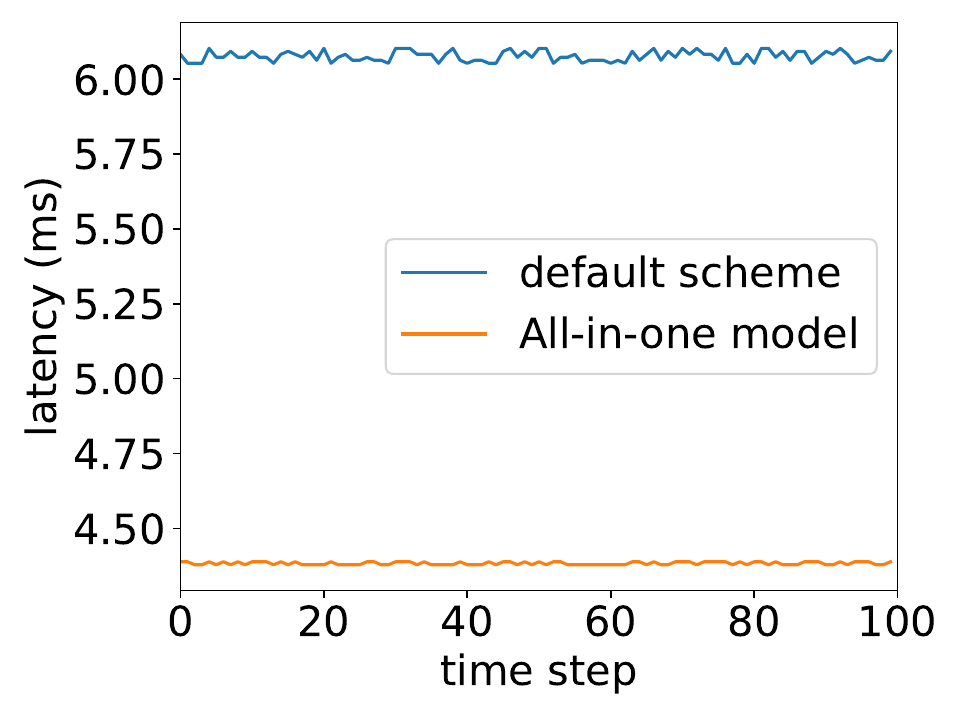}
         \caption{Latency of WCT.}
         \label{fig:aio-test-wc-l}
     \end{subfigure}
     \begin{subfigure}[b]{0.44\textwidth}
         \centering
         \includegraphics[width=\textwidth]{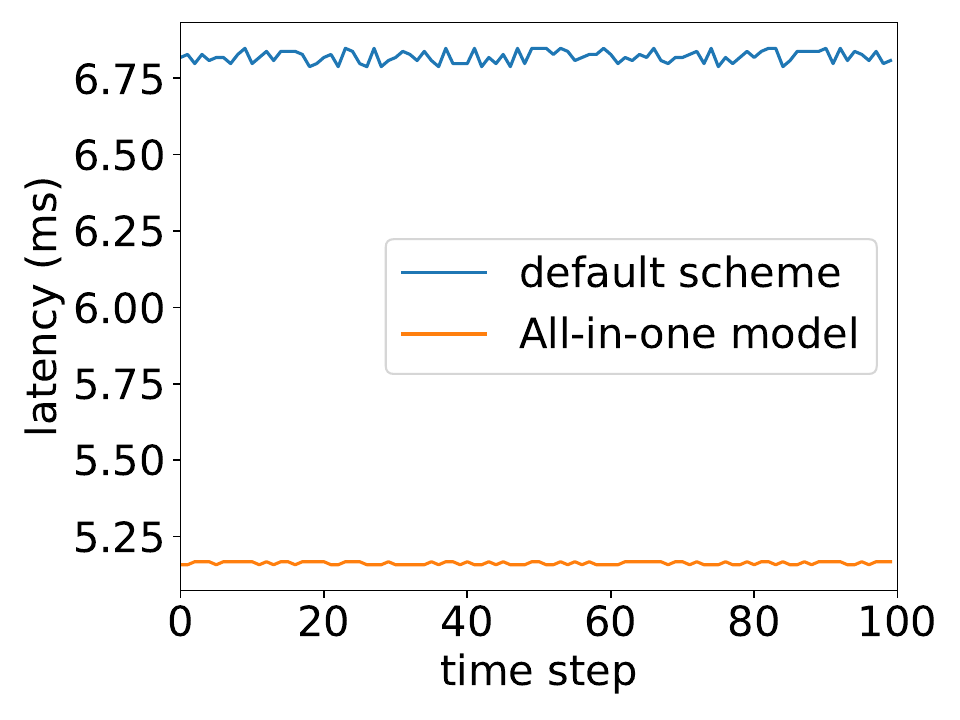}
         \caption{Latency of LSPT.}
         \label{fig:aio-test-lsp-l}
     \end{subfigure}
     \begin{subfigure}[b]{0.44\textwidth}
         \centering
         \includegraphics[width=\textwidth]{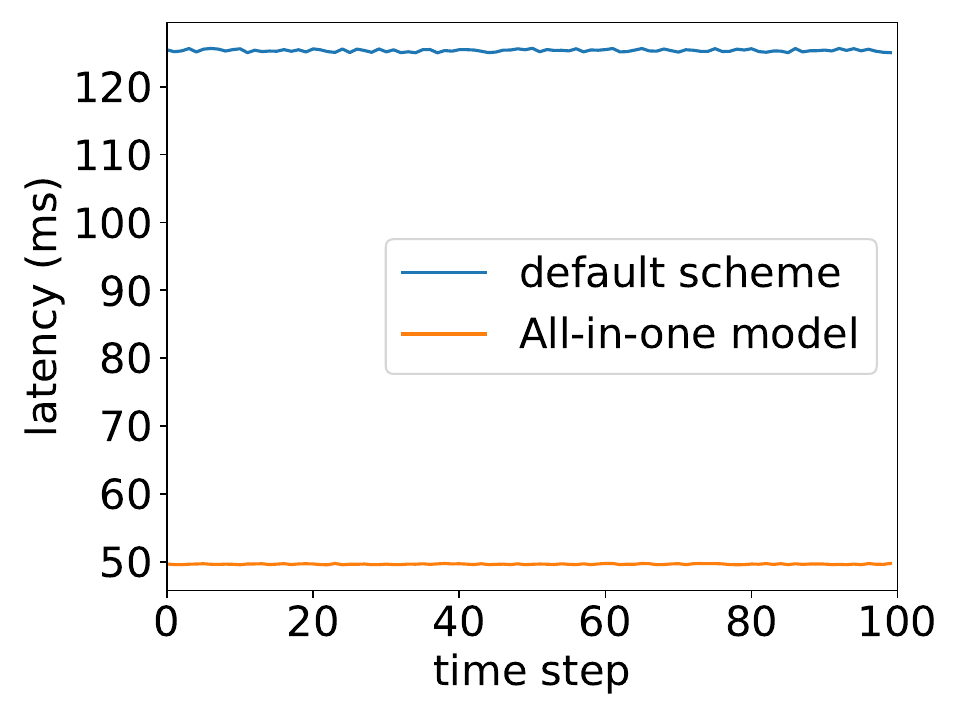}
         \caption{Latency of RGT.}
         \label{fig:aio-test-rg-l}
     \end{subfigure}
        \caption{Testing performance three topologies in a single DRL agent.}
        \label{fig:aio}
\end{figure*}

Fig.~\ref{fig:aio} shows the performance of our proposed approach compared with the system's baseline, which is labelled as the default scheme (i.e., the emission rate is without any limitation). Fig.~\ref{fig:aio-test-wc}, Fig.~\ref{fig:aio-test-lsp} and Fig.~\ref{fig:aio-test-rg} compare the throughput. Specifically, our approach provides up to 8.14\%, 4.29\% and 13.5\% higher throughput than the baseline in WCT, LSPT and RGT, respectively. It also shows superior performance on the latency reduction, demonstrating 30\%, 26\% and 13\% less than the baseline scheme.

It is noticeable that the improvement in RGT is more significant than the other two topologies because there are two branches to process data. This means one branch enters the back pressure state, the \emph{source} is suspended and does not emit data to another branch regardless of whether it is in overloading status. For example, the non-overloaded branch becomes idle with no incoming tuples during the back pressure time of another branch, which largely reduces the processed tuples by \emph{sink}. Consequently, preventing the system (i.e., both branches) from being overloaded may improve the throughput because both branches continuously process data. Moreover, the overloaded system causes a growing queue length, which leads to a significant increase in queuing latency. This can be further magnified in a topology with multiple layers, such as RGT.

Note that the result of our approach shows better stability than the baseline scheme (i.e., lighter fluctuation due to randomness). This is mainly because the baseline unstablise the system, such as growing queuing latency, exchanging back pressure messages and constantly checking back pressure status, which wastes the system's resources. It indicates that proactively avoiding overloading the system is needed.

%% file: assets/documents/06_Conclusion.tex
\section{Conclusion}
\label{Conclusion}

In this paper, a generic graph deep reinforcement learning-based rate control approach is presented to proactively throttle the data emission rate at which the data is generated to avoid overloading the system. We use graph neural networks to automatically and dynamically extract key features from the collected system metrics and forward them to actor and critic networks. The proposed method demonstrates excellent adaptability for multiple stream applications in multiple scenarios. Testing experimental results show that our approach achieves superior throughput with a reduction in the end-to-end latency.